\documentclass[showpacs,preprintnumbers,twocolumn,superscriptaddress,aps]{revtex4}
\usepackage{amssymb}
\usepackage[centertags]{amsmath}
\usepackage{txfonts}
\usepackage{epsfig}
\usepackage{bm}
\usepackage{color}
\usepackage{graphicx,graphics}
\usepackage{multirow}

\begin{document}

\title{Hadron yield correlation and constituent quark degree of freedom in heavy ion collisions}

\author{Rui-qin Wang}
\affiliation{Department of Physics, Qufu Normal University, Shandong 273165, People's
Republic of China}

\author{Feng-lan Shao}
\email{shaofl@mail.sdu.edu.cn} \affiliation{Department of Physics, Qufu Normal
University, Shandong 273165, People's Republic of China}

\author{Jun Song}
\affiliation{Department of Physics, Shandong University, Shandong 250100,
People's Republic of China}

\author{Qu-bing Xie}
\affiliation{Department of Physics, Shandong University, Shandong
250100, People's Republic of China}

\begin{abstract}

Based on the assumption of the production of deconfined quark matter, we use a
quark combination model to systematically investigate hadron yields in heavy
ion collisions from RHIC $\sqrt{s_{NN}}=200, 130, 62.4$ GeV to SPS
$E_{beam}=158, 80, 40, 30, 20 $ AGeV. We find that as the collision energy is
greater than or equal to 30 AGeV the yields of various hadrons, their
correlations, in particular, the observables
$A=\frac{\overline{\Lambda}~k^{-}~p}{\Lambda~k^{+}~\overline{p}}$ and
$B=\frac{\Lambda~k^{-}~\overline{\Xi}^{+}}{\overline{\Lambda}~k^{+}~\Xi^{-}}$ ,
are all reproduced; however, as the collision energy drops to 20 AGeV quark
combination fails. This indicates that the constituent quark degrees of freedom
represent a decisive factor in thermal hadron production above 30 AGeV and seem
to be invalid at 20 AGeV. In addition, hadron yields as well as particle ratios
at midrapidity in the most central Pb+Pb collisions at $\sqrt{s_{NN}}=5.5$ TeV
are predicted.
\end{abstract}

 \maketitle

\section{INTRODUCTION}

The question at which collision energy in heavy ion collisions the
deconfinement is first reached has attracted more and more attentions in recent
years
\cite{Andronic2009PLB,YGMa2009nuclex,Hove1982PLB,Gazdzicki,Gazdz,GAGo99,Bugaev2003PLB,Akkelin}.
The Beam Energy Scan programme of NA49 experiment at the CERN-SPS has suggested
a preliminary answer
--- around 30 AGeV\cite{20A30A2008PRC}. The ongoing Beam Energy Scan programme of STAR
Collaboration at Brookhaven National Lab provides an opportunity to study it in
more detail\cite{starESC,besRHIC}. Once the deconfined hot and dense quark
matter is produced in heavy ion collisions, the observables of various thermal
hadrons after hadronization, e.g. yields and momentum spectra etc, have some
correlations originated from early quark degrees of freedom. One of the most
typical examples is the elliptic flow (v2) of hadrons measured at RHIC
energies. As both v2 and transverse momentum($p_T$) are divided by the
constituent quark number of hadron, the rescaled v2 of various baryons and
mesons, which are just that of constituent quarks, almost coincide with each
other in the intermediate $p_T$ range \cite{AAdare2007PRL}. If the hot and
dense quark matter is hadronized by quark
(re-)combination/coalescence\cite{Fries2003PRL,Greco2003PRL,FLShao2005PRC}, as
is commonly accepted, these correlations of hadrons can be beautifully
explained. In quark (re-)combination/coalescence scenario, quarks and
antiquarks are available in unbound state before hadronization and they can
coalesce freely into various hadrons, and thereby these correlations from early
quark degrees of freedom among different hadron species are naturally formed.
On the other hand, if the deconfined quark matter is not produced at all in
collisions, there is no free quarks and antiquarks (much less their subsequent
combination) and these so-called ``quark-level" correlations of hadrons maybe
disappear or contort. Therefore, we can study whether the deconfinement is
achieved by investigating these correlations among various hadrons produced in
heavy ion collisions.

Hadron yield is one of the most significant observables from which one can
obtain a lot of important information on the hot nuclear matter produced at the
early stage of relativistic heavy ion collisions. At the high collision
energies RHIC and top SPS, the yields of different hadron species have shown an
explicit ``quark-level" correlation in quark (re-)combination/coalescence
scenario \cite{A.Bialas1998PLB,J.Zimanyi2000PLB,FLShao2005PRC}. In this paper,
we make an energy scan from RHIC energies $\sqrt{s_{NN}}=200, 130, 62.4$ GeV to
SPS energies $E_{beam}=158, 80, 40, 30, 20 $ AGeV to study at which collision
energy this ``quark-level" correlation of hadron yields first breaks. In
particular, we define two correlation quantities
$A=\frac{\overline{\Lambda}~k^{-}~p}{\Lambda~k^{+}~\overline{p}}$ and
$B=\frac{\Lambda~k^{-}~\overline{\Xi}^{+}}{\overline{\Lambda}~k^{+}~\Xi^{-}}$,
sensitive to quark degrees of freedom. The values of A and B are equal to one
in the framework of quark (re-)combination/coalesce, independent of models. The
deviation of A and B from one or not can be regarded as a possible signal of
deconfinement in heavy ion collisions. We apply a quark combination model,
which can exclusively describe hadron production and well reproduce the yields
and momentum spectra of final-state hadrons in relativistic heavy ion
collisions\cite{FLShao2007PRC,CEShao2009PRC,JSong2009MPA,DMWei2008MPA,YFWang2008CPC,WHan2009PRC},
to carry out the concrete calculations.

The paper is organized as follows. In Sec.II, we present the relations among
yields of various hadrons in quark (re-)comibination/coalescence scenario. In
Sec.III, we use the quark combination model to calculate the yields of various
hadrons, their yield ratios and correlation quantities
$A=\frac{\overline{\Lambda}~k^{-}~p}{\Lambda~k^{+}~\overline{p}}$ and
$B=\frac{\Lambda~k^{-}~\overline{\Xi}^{+}}{\overline{\Lambda}~k^{+}~\Xi^{-}}$
at midrapidity in the most central A+A collisions at different energies. Sec.IV
summaries our work.

\section{hadron yields in quark (re-)combination/coalescence scenario}
Let us start from the general inclusive formula of hadron production in quark
(re-)combination/coalescence scenario
\begin{eqnarray}
  N_M&=&\int  dp_1 dp_2 F_{q\bar{q}}(p_1,p_2)\, \mathcal {R}_M(p_1,p_2)\\
 N_B&=&\int dp_1 dp_2 dp_3 F_{qqq}(p_1,p_2,p_3)\, \mathcal {R}_B(p_1,p_2,p_3).
\end{eqnarray}
Here $F_{q\bar{q}}$ ($F_{qqq}$) is the joint quark-antiquark (three quark)
distribution. $\mathcal {R}_M$ ($\mathcal {R}_B$) is the combination function
which stands for the formation probability of quark antiquark (three quarks)
into a meson (baryon), dominated by chromodynamics. In sudden approximation, it
is equal to the overlap between two (three) quark wave functions and the wave
function of meson (baryon). Neglecting exotic (multi-quark) states, mesons and
baryons exhaust all fate of quarks and antiquarks. One reaches the following
relations : $\sum N_{M}+3\sum N_{B}=\sum N_q$ and $\sum N_{M}+3\sum
N_{\bar{B}}=\sum N_{\bar{q}}$, where $N_{q}$ ($N_{\bar{q}}$) is the quark
(antiquark) number of flavor $q$($\bar{q}$). Extracting $N_{q}$ and
$N_{\bar{q}}$ from the joint two (three) quark distribution $F_{q\bar{q}}$
($F_{qqq}$) and putting them out of the integral, one has the following
schematic relations between hadron yields and quark numbers after integrating
over quark momenta
\begin{equation}
  N_{M(q\bar q)}\propto C_{M}N_{q}N_{\bar q} , \hspace{1cm}
  N_{B({qqq})}\propto C_{B}N_{q}N_{q}N_{q},
\end{equation}
and quark number conservation will fix the proportionality coefficient. The
effects of combination function on hadron yields are characterized with the
factors $C_{M}$ and $C_{B}$, and $C_{M}=C_{\overline{M}}$ and
$C_{B}=C_{\overline{B}}$ are assumed.

One realized method of quark number conservation during combination is adding a
factor $b_q$ for each quark flavor in above equations, as did in ALCOR model
\cite{alcor95},
\begin{eqnarray}
  N_{M(q\bar q)}&=& C_{M}(b_{q}N_{q})(b_{\bar q}N_{\bar q}),\label{mcom}\\
  N_{B({qqq})}&=& C_{B}(b_{q}N_{q})(b_{q}N_{q})(b_{q}N_{q}).\label{bcom}
\end{eqnarray}
Then $b_{q}$ can be uniquely determined by quark number conservation. According
to Eqs.(\ref{mcom}) and (\ref{bcom}), we have the following relations between
hadrons and the corresponding antihadrons
\begin{eqnarray}
\frac{\overline{p}}{p}&=&(\frac{b_{\overline{q}}\overline{q}}{b_{q}q})^{3},
\hspace{1.6cm}
\frac{k^{+}}{k^{-}}=(\frac{b_{q}q}{b_{\overline{q}}\overline{q}})(\frac{b_{\overline{s}}\overline{s}}{b_{s}s}),
\nonumber \\
\frac{\overline{\Lambda}}{\Lambda}&=&(\frac{b_{\overline{q}}\overline{q}}{b_{q}q})^{2}\frac{b_{\overline{s}}\overline{s}}{b_{s}s},
\hspace{0.9cm}
\frac{\Xi^{-}}{\overline{\Xi}^{+}}=(\frac{b_{s}s}{b_{\overline{s}}\overline{s}})^{2}\frac{b_{q}q}{b_{\overline{q}}\overline{q}}.
\nonumber
\end{eqnarray}
Here, we use particle symbols stand for their numbers for short, $q$ for light
quark number and $s$ for strange quark number.  Hiding the quark content in
hadron yield, we obtain the following interesting relations among different
hadron species
\begin{equation}
\frac{\overline{\Lambda}~k^{-}}{\Lambda~k^{+}}=\frac{\overline{p}}{p},
\hspace{1cm}
\frac{\Lambda~k^{-}}{\overline{\Lambda}~k^{+}}=\frac{\Xi^{-}}{\overline{\Xi}^{+}}.
\nonumber
\end{equation}
We define correlation quantities $A$ and $B$ as follows
\begin{equation}
 A=\frac{\overline{\Lambda}~k^{-}~p}{\Lambda~k^{+}~\overline{p}} \hspace{1cm}
B=\frac{\Lambda~k^{-}~\overline{\Xi}^{+}}{\overline{\Lambda}~k^{+}~\Xi^{-}}.
\end{equation}
If the quark matter exists and hadronizes via sudden (re-)combination/coalesce
in heavy ion collisions, A and B should be equal to one for directly produced
hadrons. It is a general result under the constraint of quark number
conservation, which is independent of specific models.

Different from the well-known recombination and coalescence models
\cite{Fries2003PRL,Greco2003PRL}, the quark combination model
\cite{FLShao2005PRC,QBXie1988PRD} is unique for its combination rule. The main
idea of the combination rule is to line up the (anti)quarks in a
one-dimensional order in phase space, e.g., in rapidity, and then let them
combine into initial hadrons one by one according to this
order\cite{FLShao2005PRC}. Three (anti)quarks or a quark-antiquark pair in the
neighborhood form a (anti)baryon or a meson, respectively. At last all quarks
and antiquarks are combined into hadrons. The relations between hadron yields
and the corresponding quark numbers are easily obtained \cite{CEShao2009PRC}.
With this rule, the model can give the yields and momentum distributions of all
hadrons (included in the model) in an event, possessing some exclusive nature.
The decay of the short-life resonances is systematically taken into account in
the model. The model has been realized in Monte Carlo program and has described
many properties of hadron production in relativistic heavy ion collisions
\cite{FLShao2007PRC,CEShao2009PRC,JSong2009MPA,DMWei2008MPA,YFWang2008CPC,WHan2009PRC}.

\section{RESULTS AND DISCUSSIONS}

In this section, we use the quark combination model to calculate the hadron
yields and their ratios as well as correlation quantities
$A=\frac{\overline{\Lambda}~k^{-}~p}{\Lambda~k^{+}~\overline{p}}$ and
$B=\frac{\Lambda~k^{-}~\overline{\Xi}^{+}}{\overline{\Lambda}~k^{+}~\Xi^{-}}$
at midrapidity in the most central A+A collisions at
$\sqrt{s_{NN}}=200,~130,~62.4$ GeV and $E_{beam}=158,~80,~40,~30,~20$ AGeV. The
predictions at LHC are also presented. The necessary input of the model, i.e.,
quark distribution just before hadronization, can be obtained by applying the
hydrodynamics to describe the evolution of hot and dense quark matter just
before hadronization, see Appendix \ref{app1} for details.

\subsection{Hadron yields, their ratios and correlation quantities A and B}

\begin{table*}[!htbp]
\caption{The calculated hadron yields $dN/dy$ at midrapidity in the
most central A+A collisions at different energies. The experimental
data are taken from
Refs\cite{20A30A2008PRC,20A30A40A80A158A2006PRC,20A30A40A80A158A2008PRC,
40A80A158A2004PRL,40A158A2005PRL,158A2002PRC,622009nuclex,1302002PRL,1302004PRL,1302004PRC,2002004PRC,2002007PRL}.}
\label{yield}
{\begin{tabular*}{170mm}{@{\extracolsep{\fill}}c|c|c|c|c|c|c|c|c}
\toprule
&\multicolumn{2}{c|}{Au+Au~200 GeV} &\multicolumn{2}{c|}{Au+Au~130 GeV}&\multicolumn{2}{c|}{Au+Au~62.4 GeV}&\multicolumn{2}{c}{Pb+Pb~158 AGeV}\\
\colrule
&data& model &data& model &data& model &data& model \\
\colrule
$\pi^{+}$ &$286.4\pm24.2$ &281.0 &$276\pm3\pm35.9$ &267.2 &$233\pm17$&227.4  &$170.1\pm0.7\pm9$&165.1 \\
$\pi^{-}$ &$281.8\pm22.8$ &281.8 &$270\pm3.5\pm35.1$&270.5 &$237\pm17$&233.5 &$175.4\pm0.7\pm9$&176.5\\
\colrule
$k^{+}$ &$48.9\pm6.3$&48.6 &$46.7\pm1.5\pm7.0$&45.2 &$37.6\pm2.7$&38.3 &$29.6\pm0.3\pm1.5$&27.2\\
$k^{-}$ &$45.7\pm5.2$&46.1 &$40.5\pm2.3\pm6.1$&42.4 &$32.4\pm2.3$&32.2 &$16.8\pm0.2\pm0.8$&16.2\\
\colrule
$p$ &$18.4\pm2.6$&17.0 &$28.7\pm0.9\pm4.0$&25.7 &$29.0\pm3.8$&29.1 &$29.6\pm0.9\pm2.9$&30.1\\
$\overline{p}$ &$13.5\pm1.8$&12.5 &$20.1\pm1.0\pm2.8$&18.2 &$13.6\pm1.7$&13.5 &$1.66\pm0.17\pm0.16$&1.91\\
\colrule
$\Lambda$ &$16.7\pm0.2\pm1.1$&15.3 &$17.3\pm1.8\pm2.8$&14.5 &$14.9\pm0.2\pm1.49$&13.7 &$10.9\pm1.0\pm1.3$&13.3\\
$\overline{\Lambda}$ &$12.7\pm0.2\pm0.9$&12.1 &$12.7\pm1.8\pm2.0$&10.9 &$8.02\pm0.11\pm0.8$&7.28 &$1.62\pm0.16\pm0.2$&1.66\\
\colrule
$\Xi^{-}$ &$2.17\pm0.06\pm0.19$&2.05 &$2.04\pm0.14\pm0.2$&1.93 &$1.64\pm0.03\pm0.014$&1.67 &$1.44\pm0.10\pm0.15$&1.18\\
$\overline{\Xi}^{+}$ &$1.83\pm0.05\pm0.20$&1.69 &$1.74\pm0.12\pm0.17$&1.53 &$0.989\pm0.057\pm0.057$&1.02 &$0.31\pm0.03\pm0.03$&0.23\\
\colrule
$\Omega^{-}$ & & & & & & &$0.14\pm0.03\pm0.01$&0.11\\
$\overline{\Omega}^{+}$
&\raisebox{1.6ex}[0pt]{$0.53\pm0.04\pm0.04$}&\raisebox{1.6ex}[0pt]{0.56}
&\raisebox{1.6ex}[0pt]{$0.56\pm0.11\pm0.06$}&\raisebox{1.6ex}[0pt]{0.52}
&\raisebox{1.6ex}[0pt]{$0.356\pm0.046\pm0.014$}&\raisebox{1.6ex}[0pt]{0.379}
&$0.07\pm0.02\pm0.01$&0.04\\
\colrule
$\chi^{2}/ndf$ &\multicolumn{2}{c|}{$2.8/8$} &\multicolumn{2}{c|}{$2.0/8$} &\multicolumn{2}{c|}{$2.1/8$} &\multicolumn{2}{c}{$8.5/9$}\\
\end{tabular*}}

\begin{tabular*}{170mm}{@{\extracolsep{\fill}}c|c|c|c|c|c|c|c|c}
\toprule
&\multicolumn{2}{c|}{Pb+Pb~80 AGeV} &\multicolumn{2}{c|}{Pb+Pb~40 AGeV} &\multicolumn{2}{c|}{Pb+Pb~30 AGeV} &\multicolumn{2}{c}{Pb+Pb~~20 AGeV}\\
\colrule
&data&model &data&model &data&model &data&model \\
\colrule
$\pi^{+}$ &$132.0\pm0.5\pm7$&129.9 &$96.6\pm0.4\pm6$&97.8 &$83.0\pm0.4\pm4.2$&84.8 &$72.9\pm0.3\pm3.6$&73.6\\
$\pi^{-}$ &$140.4\pm0.5\pm7$&141.8 &$106.1\pm0.4\pm6$&110.1 &$96.5\pm0.5\pm4.8$&99.3 &$84.8\pm0.4\pm4.2$&85.0\\
\colrule
$k^{+}$ &$24.6\pm0.2\pm1.2$&23.8 &$20.1\pm0.3\pm1.0$&19.4 &$21.2\pm0.8^{+1.5}_{-0.9}$&21.6 &$16.4\pm0.6\pm0.4$&16.6 \\
$k^{-}$ &$11.7\pm0.1\pm0.6$&11.9 &$7.58\pm0.12\pm0.4$&7.27 &$7.8\pm0.1\pm0.2$&7.3 &$5.58\pm0.07\pm0.11$&5.01 \\
\colrule
$p$ &$30.1\pm1.0\pm3.0$&31.5 &$41.3\pm1.1\pm4.1$&38.6 &$42.1\pm2.0\pm4.2$&37.7 &$46.1\pm2.1\pm4.6$&37.0 \\
$\overline{p}$ &$0.87\pm0.07\pm0.09$&0.83 &$0.32\pm0.03\pm0.03$&0.36 &$0.16\pm0.02\pm0.02$&0.17 &$0.06\pm0.01\pm0.006$&0.07 \\
\colrule
$\Lambda$ &$13.5\pm0.7\pm1.0$&14.5 &$15.3\pm0.6\pm1.0$&14.7 &$14.7\pm0.2\pm1.2$&14.2 &$13.4\pm0.1\pm1.1$&11.5\\
$\overline{\Lambda}$ &$1.06\pm0.08\pm0.1$&0.95 &$0.42\pm0.04\pm0.04$&0.49 &$0.21\pm0.02\pm0.02$&0.20 &$0.10\pm0.02\pm0.01$&0.08 \\
\colrule
$\Xi^{-}$ &$1.22\pm0.14\pm0.13$&1.23 &$1.15\pm0.11\pm0.13$&1.10 &$1.17\pm0.13\pm0.13$&1.38 &$0.93\pm0.13\pm0.10$&0.92\\
$\overline{\Xi}^{+}$ &$0.21\pm0.03\pm0.02$&0.15 &$0.07\pm0.01\pm0.01$&0.08 &$0.05\pm0.01\pm0.01$&0.07 &$------$&0.02\\
\colrule
$\Omega^{-}+\overline{\Omega}^{+}$ &$------$&0.13 &$0.10\pm0.02\pm0.02$&0.09 &$------$&0.14 &$------$&0.07\\
\colrule
$\chi^{2}/ndf$ &\multicolumn{2}{c|}{$2.9/7$} &\multicolumn{2}{c|}{$3.0/8$} &\multicolumn{2}{c|}{$5.6/7$} &\multicolumn{2}{c}{$15.3/6$}\\
\botrule
\end{tabular*}
\end{table*}

Table \ref{yield} shows the hadron density dN/dy at midrapidity in the most
central A+A collisions at RHIC energies 200, 130, 62.4 GeV and SPS energies
158, 80, 40, 30, 20 AGeV. The experimental data are taken from
Refs\cite{20A30A2008PRC,20A30A40A80A158A2006PRC,20A30A40A80A158A2008PRC,
40A80A158A2004PRL,40A158A2005PRL,158A2002PRC,622009nuclex,1302002PRL,1302004PRL,1302004PRC,2002004PRC,2002007PRL}.
The agreement between the calculation results and the experimental data is good
except at 20 AGeV where $\chi^{2}/ndf$ is far greater than one. We note that
ALCOR model \cite{alcor95} can also describe well the hadron yields at top SPS
energy, which gives a cross-verification of quark combination. On the contrary,
HIJING or HIJING/B model which is in compliance with the hadronic scenario for
the early evolution of heavy ion collisions via a fragmentation hadronization,
can not self-consistently explain the data of multi-strange hadrons
\cite{topor94,Csizmadia99}.

\begin{figure}[!htb]
  \includegraphics[width=\linewidth]{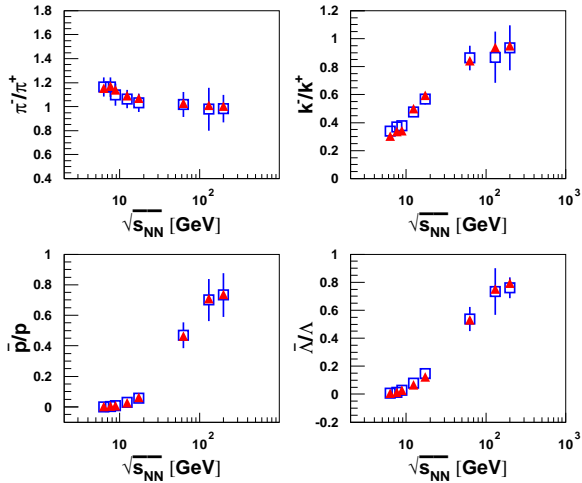}\\
  \caption{(Color online)The yield ratios of antihadrons to hadrons at midrapidity as the function of the collision energy.
The filled symbols are the calculated results, and the experimental data, open
symbols with error bar, are from
Refs\cite{20A30A2008PRC,20A30A40A80A158A2006PRC,20A30A40A80A158A2008PRC,
40A80A158A2004PRL,158A2002PRC,622009nuclex,1302002PRL,1302004PRC,2002004PRC,2002007PRL}.}\label{anhiration}
\end{figure}

Fig.\,\ref{anhiration} shows the ratios of antihadrons to hadrons at midrapidity as the
function of collision energy. The filled triangles are the calculated results, and the
experimental data are taken from
Refs\cite{20A30A2008PRC,20A30A40A80A158A2006PRC,20A30A40A80A158A2008PRC,
40A80A158A2004PRL,158A2002PRC,622009nuclex,1302002PRL,1302004PRC,2002004PRC,2002007PRL}.
These ratios are mainly influenced by the no-zero baryon number density. As the collision
energy increases, the nuclear transparency power becomes strong and baryon number density
at midrapidity becomes small. This results in a rapid increase for ratios of
$K^{-}/K^{+}$, $\overline{p}/p$, $\overline{\Lambda}/\Lambda$ with the increasing
collision energy. $\pi^{-}/\pi^{+}$ follows a different pattern. At low collision
energies $\pi^{-}/\pi^{+}$ is slightly higher than one, while it is close to one at high
energies. This is caused by the asymmetry of decay contribution from hyperons and
anti-hyperons (e.g. $\Lambda\rightarrow p\, \pi^{-}$). As the collision energy increases,
the yields of hyperons are close to that of anti-hyperons and their decay contributions
to pion yields are almost the same and therefore the ratio of $\pi^{-}/\pi^{+}$ is close
to one.

\begin{figure}[!htb]
  \includegraphics[width=1.0\linewidth]{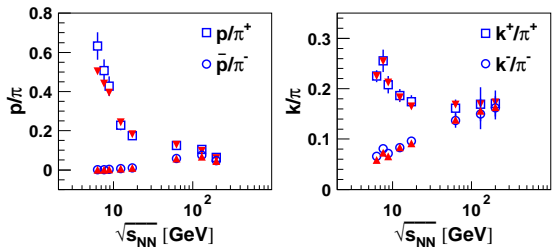}\\
  \caption{(Color online)The values of the relative production yields $p/\pi$ and $k/\pi$ at different energies. The
 triangles down are the numerical results for $p/\pi^+$ and $k^+/\pi^+$, and the triangles
up are the numerical results for $\overline{p}/\pi^-$ and $k^-/\pi^-$. The experimental
data, open symbols with error bar, are from
Refs\cite{20A30A2008PRC,20A30A40A80A158A2006PRC,
158A2002PRC,622009nuclex,1302002PRL,1302004PRC,2002004PRC}.}\label{ppiratio}
\end{figure}

Fig.\,\ref{ppiratio} shows the ratios of baryons to mesons $p/\pi$ and the strangeness
ratios $K/\pi$ at midrapidity as the function of collision energy. The filled triangles
(up and down) are the computed results, and the experimental data are taken from
Refs\cite{20A30A2008PRC,20A30A40A80A158A2006PRC,
158A2002PRC,622009nuclex,1302002PRL,1302004PRC,2002004PRC}. The big splits between
$p/\pi^{+}$ and $\overline{p}/\pi^{-}$ and between $K^{+}/\pi^{+}$ and $K^{-}/\pi^{-}$ at
low collision energies is due to the high baryon number density. At 20 AGeV, the result
of $p/\pi^{+}$ deviates seriously from the data. This is probably because the participant
nucleons are not broken completely in collisions. These nucleon fragments deposited in
midrapidity region lead to the extra contribution of proton production besides those from
quark combination and lead to the excessively high ratio of $p/\pi^+$.

\begin{figure}[!htbp]
  \includegraphics[width=\linewidth]{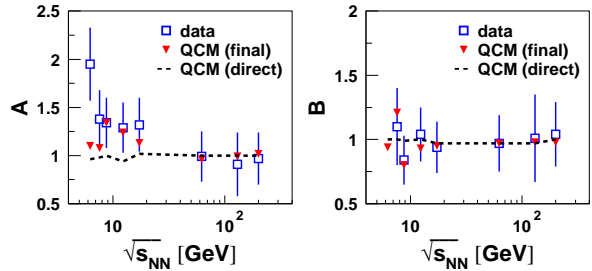}\\
  \caption{(Color online)The correlation quantities A and B as the function of collision energy.
   The experimental data, open symbols with error bar, are from Refs\cite{20A30A2008PRC,20A30A40A80A158A2006PRC,20A30A40A80A158A2008PRC,
40A80A158A2004PRL,158A2002PRC,622009nuclex,1302002PRL,1302004PRL,1302004PRC,2002004PRC,2002007PRL}.}
  \label{ab}
\end{figure}

Following the experimental data in Table \ref{yield}, A and B at different
collision energies are evaluated and the results are presented with squares in
Fig.\ref{ab}. We find that the data of A and B at RHIC energies are almost
equal to one while at SPS energies they deviate from one; particularly, at 20
AGeV the value of A amounts to two, seriously deviating from one (the data for
B are unavailable). This is probably because the decay of resonances will blur
A and B to a certain degree. In order to explore the decay effect, we use the
quark combination model to compute the values of A and B for the directly
produced hadrons and the final-state hadrons, respectively. The dashed lines
are the results of directly produced hadrons. Just as analyzed above, the
dashed lines keep an invariant value of one for both A and B, independent of
collision energy. The very small fluctuations of dashed lines are due to the
fact that the rapidity distributions of the formed hadrons are slightly
different from those of quarks, which leads to a small amount of hadrons formed
by midrapidity quarks escape from the midrapidity region \cite{FLShao2007PRC}.
The filled triangles down are model results of final-state hadrons. We see that
as collision energy is greater than 30 AGeV these triangles agree well with the
experimental data within statistical uncertainties. Removing the part of
resonance decay, the data of A and B for directly produced hadrons should be
equal to one, respectively. This suggests the existence of the ``quark level"
correlation of directly produced hadrons at these collision energies. However,
at 20 AGeV the value of A calculated via quark combination for final state
hadrons  seriously deviates the data. This in itself indicates the quark
degrees of freedom do not represent a decisive factor in hadron production at
20 AGeV.

From the above analysis of hadron yields, hadron ratios and correlation
quantities A and B, we find that as the collision energy is greater than or
equal to 30 AGeV the quark combination can describe reasonably well all these
observables which indicates the existence of constituent quark degrees of
freedom in the energy region. When the collision energy drops to 20 AGeV,
however, the quark combination mechanism can not self-consistently describe
these quantities, in particular the correlation quantity A. Note that in order
to best fit the data of hadron yields, we have adjusted the strangeness
parameter $\lambda_s$ (see Table \ref{inputs}) to the quite high values at 20,
30 AGeV which are far greater than the saturated values at higher collision
energies and are also greater than those at lower energies
\cite{thermalreview}.  But even so the quark combination can not
self-consistently explain the data at 20 AGeV. In addition, the high
strangeness entangled with baryon density leads to the Sawtooth-like shape for
calculated final-state A and B at low SPS energies. In fact this peak behavior
of strangeness (i.e. $K^{+}/\pi^{+}$ ratio or $\lambda_s$ determined mainly by
the former ) at low SPS energies has been interpreted in Ref
\cite{Gazdzicki,GAGo99} as a result of onset of deconfinement, i.e. the result
of strangeness carrier changing from strange hadrons to strange quarks at the
onset of deconfinement. The failure of quark combination energy indicates the
(partial) disappearance of constituent quark degrees of freedom and closely
relates to the onset of deconfinement around 30 AGeV observed by NA49
Collaboration \cite{20A30A2008PRC}.

\subsection{Predictions at LHC}
With the increasing collision energy, the strangeness $\lambda_s$ of the hot
and dense quark matter in Table \ref{inputs} tends to be saturated and the
squared sound velocity $c_s^2$ approaches an ideal value $1/3$ and the baryon
number density $n_{0}$ decreases regularly. So we take $\lambda_s$=0.43,
$c_s^2=1/3$ and $n_{0}=0.0 fm^{-3}$ at $\sqrt{s_{NN}}=5.5$ TeV.  The initial
entropy density of the central point at the beginning of hydrodynamic evolution
is taken to be 256.5 /fm$^3$ according to Eq. (\ref{initEn}). In Tables
\ref{yeild_LHC} and \ref{ratio_LHC}, we present our predictions of hadron
yields and hadron ratios at LHC energy ($\sqrt{s_{NN}}=5.5$ TeV). Our computed
particle ratios are consistent with the results of the thermal model in
Ref\cite{Andronic2009PLB}.

\begin{table}[!htbp]
\caption{The predicted $dN/dy$ of identified hadrons at midrapidity in the most
central ($0-5\%$) Pb+Pb collisions at $\sqrt{s_{NN}}=5.5$ TeV.}
{\begin{tabular}{@{}cccccccccccc@{}} \toprule $\pi^{+}$ &$\pi^{-}$ &$k^{+}$
&$k^{-}$ &$p$ &$\overline{p}$ &$\Lambda$ &$\overline{\Lambda}$ &$\Xi^{-}$
&$\overline{\Xi}^{+}$ &$\Omega^{-}$
&$\overline{\Omega}^{+}$\\
\colrule
435.5 &435.5 &71.6 &71.6 &34.6 &34.6 &20.5 &20.5 &2.9 &2.9 &0.4 &0.4\\
\botrule
\end{tabular} \label{yeild_LHC}}
\end{table}

\begin{table}[!htbp]
\caption{The predicted hadron ratios at midrapidity in the most
central Pb+Pb collisions at $\sqrt{s_{NN}}=5.5$ TeV.}
 {\begin{tabular}{@{}ccccccc@{}} \toprule
$p/\pi^{+}$ &$\overline{p}/\pi^{-}$ &$k^{+}/\pi^{+}$ &$k^{-}/\pi^{-}$
&$\Lambda/\pi^{-}$ &$\Xi^{-}/\pi^{-}$ &$\Omega^{-}/\pi^{-}$\\
\colrule
0.079 &0.079 &0.165 &0.165 &0.047 &0.007 &0.001\\
\botrule
\end{tabular} \label{ratio_LHC}}
\end{table}

\section{Summary}

In this paper, we used the quark combination model to make a systematical study
of the hadron yields in heavy ion collision in a broad collision energy region.
At the collision energies where the deconfined quark matter has been created,
the yields of various hadrons after hadronization have some correlations
inherited from the early quark degrees of freedom. We investigate at which
collision energy this ``quark-level" correlation of hadron yields first breaks.
We apply the hydrodynamics to describe the evolution of deconfined quark matter
and to obtain the quark distribution just before hadronization, then utilize
the quark combination model to describe hadronization. We find that as the
collision energy is greater than or equal to 30 AGeV, the quark combination
well reproduce the yields of various hadrons, their ratios and correlation
quantities A and B; however, as the collision energy drops to 20 AGeV, the
mechanism can not self-consistently describe these quantities. This indicates
that the constituent quark degrees of freedom represent a decisive factor in
thermal hadron production above 30 AGeV and seem to be invalid at 20 AGeV. It
is related to the onset of deconfinement observed at collision energy 20-30
AGeV. Finally, we predict the yields of various hadrons and their ratios in the
most central Pb+Pb collisions at $\sqrt{s_{NN}}=5.5$ TeV.
\subsection*{ACKNOWLEDGMENTS}
The authors thank Z. T. Liang, Q. Wang and G. Li for helpful discussions. R. Q.
Wang would like to thank L. G. Pang and J. Deng for fruitful discussions. The
work is supported in part by the National Natural Science Foundation of China
under the grant 10775089, 10947007 and 10975092.

\appendix
\section{Quark distribution just before hadronization}\label{app1}
\begin{table*}[!htb]
\caption{The values of the initial baryon density $n_{0}$, strangeness factor
$\lambda_{s}$ and squared sound velocity $c_s^{2}$ in central A+A collisions at
different energies.}
\begin{tabular}{@{}cccccccccc@{}}
\toprule Energy &200GeV &130GeV &62.4GeV &158AGeV &80AGeV &40AGeV &30AGeV &20AGeV \\
\colrule
$n_{0}$ ($fm^{-3}$)&0.30 &0.34 &0.74 &1.31 &1.46 &1.64 &1.74 &1.56 \\
$\lambda_{s}$ &0.43 &0.43 &0.43 &0.44 &0.50 &0.57 &0.80 &0.70 \\
$c_{s}^{2}$ &$1/3.1$ &$1/3.4$ &$1/4.0$ &$1/6.0$ &$1/6.0$ &$1/6.0$ &$1/6.0$ &$1/6.0$ \\
\botrule
\end{tabular} \label{inputs}
\end{table*}

The quark distribution just before hadronization is needed for the quark
combination to describe the hadron production in relativistic heavy ion
collisions. Here, we use relativistic hydrodynamics
\cite{Landau53,DirkH1995NPA,Kolb0305084nuth} to describe the time-space
evolution of the hot and dense quark matter before hadronization. The evolution
equation of hydrodynamics follows from the local conservation laws for energy,
momentum, and other conserved charges, e.g. baryon number,
\begin{eqnarray}
\partial_{\mu}T^{\mu\nu}(x)=0,~~(\nu=0,1,2,3)
\end{eqnarray}
\begin{eqnarray}
\partial_{\mu}j^\mu(x)=0,
\end{eqnarray}
by inserting the ideal fluid decomposition
\begin{eqnarray}
T^{\mu\nu}(x)=\bigg(e(x)+p(x)\bigg)u^\mu(x)u^\nu(x)-g^{\mu\nu}p(x),
\end{eqnarray}
\begin{eqnarray}
j^\mu(x)=n(x)u^\mu(x).
\end{eqnarray}
Here, $u^\mu(x)=\gamma(1,v_x,v_y,v_z)$ with
$\gamma=1/\sqrt{1-v_x^2-v_y^2-v_z^2}$ is the local four velocity of a
thermalized fluid cell; $e(x)$ is the energy density, $p(x)$ the pressure, and
$n(x)$ the conserved number density.

As the energy density of the fluid cell drops to 1.0 GeV/fm$^3$ (a common
criteria of phase transition from Lattice QCD\cite{Karsch2002NPA}), we stop the
hydrodynamic evolution and let the constituent quarks and antiquarks freeze out
according to Cooper-Frye formalism \cite{Coopfrye74}
\begin{equation}
E\frac{dN_i}{d^3p}=\frac{dN_i}{dyp_Tdp_Td\varphi}=\frac{g_i}{(2\pi)^3}\int_{\Sigma}f_i(p\cdot
u(x),x)p\cdot d^3\sigma(x), \label{coopfry}
\end{equation}
where $d^3\sigma(x)$ is the outward normal vector on the freeze-out surface
$\Sigma(x)$, $g_i$ the degeneracy factor of quarks ($g_i=6$). The phase-space
distribution $f$ in the formula is taken to be a local equilibrium
distribution,
\begin{equation}
 f_i(E,x)=\frac{1}{\exp[(E-\mu_i(x))/T(x)]+1}.
\end{equation}
We consider only the freeze-out of light and strange quarks and antiquarks. The
three chemical potentials $\mu_{q}$, $\mu_{\bar{q}}$ and
$\mu_{s}=\mu_{\bar{s}}$ at freeze-out can be determined uniquely by the global
conservation of energy and baryon number plus an ancillary constraint of
strangeness.

We study in this paper the hadron production in midrapidity region only. The
code of Kolb\cite{Kolb2000PRC,Kolb2001NPA,Kolb2003PRC} for 2+1- dimensional
hydrodynamics with the longitudinal boost invariance is used to simulate the
evolution of quark system before hadronization.

These frozen-out quarks and antiquarks with momentum distributions in
Eq.\,(\ref{coopfry}) are hadronized by the quark combination
model\cite{FLShao2005PRC,QBXie1988PRD}. In this method of hydrodynamics + quark
combination, the quark combination model is default setting while some inputs
for hydrodynamics need fixing. The first is initial entropy density $s_{0}$ of
the central point at the beginning of hydrodynamic evolution. The distribution
of transverse entropy density is determined by optical Glauber
model\cite{Glauber1959}. We apply the following empirical formula of
$s_{0}$\cite{entropy2008nuth} as the function of collision energy
$\sqrt{s_{NN}}$
\begin{equation}
s_{0}(\sqrt{s_{NN}})= 6.99\times 10^{-3}\Big(312.5\log_{10}{\sqrt{s_{NN}}} -
64.8\Big)^{3/2}. \label{initEn}
\end{equation}
The second is the initial baryon number density $n_{0}$ at central point fixed
by the data of the net-proton rapidity density. The distribution of transverse
baryon number density is also determined by optical Glauber model. The third is
the equation of state $p=p(e)$ for the hot and dense quark matter which is
taken to be the simplest pattern $p=c_s^{2}e$. The squared sound velocity
$c_s^{2}$ is obtained by fitting the data of transverse momentum spectra of
protons. The fourth is the strangeness of hot and dense quark matter denoted by
the factor $\lambda_{s}=2\langle s\bar{s}\rangle/\langle u\bar{u}+d\bar{d}
\rangle$. In order to make a better description of strange hadrons, we regard
it as a free parameter in the present paper. The values of $n_{0}$,
$\lambda_{s}$ and $c_s^{2}$ in central A+A collisions at different energies are
shown in Table \ref{inputs}. We note that the extracted $\lambda_{s}$ around 30
AGeV are quite high which are consistent with the analytic results of thermal
model \cite{thermalreview}.

\end{document}